\documentclass[aps,physrev,lengthcheck,superscriptaddress,nofootinbib,onecolumn]{revtex4-2}
\usepackage[colorlinks,allcolors=BrickRed,unicode]{hyperref}
\usepackage[dvipsnames]{xcolor}
\usepackage{amsmath,amssymb,amsfonts,physics,bigints,mathtools,amsthm}
\usepackage[version=4]{mhchem}
\usepackage{graphicx}
\usepackage{appendix}
\pdfoutput=1

\newcommand{\eq}[1]{\begin{align}#1\end{align}}
\newcommand{\bs}{\boldsymbol}
\newcommand{\mr}{\mathrm}
\newcommand{\DD}{{\mathcal D}}

\theoremstyle{definition}
\newtheorem{assumption}{Assumption}

\newtheorem{manualassumptioninner}{Assumption}
\newenvironment{manualassumption}[1]{
  \renewcommand\themanualassumptioninner{#1}
  \manualassumptioninner
}{\endmanualassumptioninner}

\begin{document}

\title{Statistical mechanics of fluids with hidden degrees of freedom}

\author{Masanari Shimada}

\email{shimada@sat.t.u-tokyo.ac.jp}

\affiliation{Institute of industrial science, the University of Tokyo}

\author{Tetsuya J. Kobayashi}

\affiliation{Institute of industrial science, the University of Tokyo}

\date{\today}

\begin{abstract}
Although coarse-grained models have been widely used to explain exotic phenomena in complex fluids, such as droplet formation in living cells, these conventional approaches often fail to capture the intricate microscopic degrees of freedom that such fluids inherently possess.
In this study, we propose a model that incorporates distinct microscopic degrees of freedom and their interactions, without directly relying on conventional coarse-grained descriptions.
By introducing two key assumptions, we show that the system can exhibit equilibrium states characterized by heterogeneous density profiles with finite length scales, resembling those typically associated with non-equilibrium phenomena.
These findings highlight the importance of distinguishing between equilibrium states and non-equilibrium steady states in highly complex systems.
\end{abstract}

\maketitle 

\renewcommand{\theequation}{\arabic{section}.\arabic{equation}}

\section{Introduction}

Complex fluids are mixtures of various types of liquids or gases, often composed of molecules with complicated internal structures~\cite{Larson1998The}.
These systems are ubiquitous in nature and are undoubtedly of great scientific and engineering importance.
From an engineering perspective, the fluids we encounter in daily life such as air, seawater, and beverages, are almost always mixtures of multiple components.
From a scientific perspective, these complex systems, which contain many distinct chemical species or, more generally, degrees of freedom (DOF), exhibit rich macroscopic behavior that is absent in simple fluids and have attracted significant attention in the field of soft condensed matter physics.

Despite their importance, our understanding of complex fluids remains very limited even for thermal equilibrium states, whereas simple fluids are relatively well understood~\cite{Hansen2013Theory}.
When a system contains a large number of DOF, it can exhibit complex phase separation, and the resulting morphologies are generally intractable.
As studied in Ref.~\cite{Sear2003Instabilities}, even with a simple Flory–Huggins free energy, the resulting liquid structures can be highly nontrivial in multi-component systems.
Furthermore, due to the complexity of equilibrium states, it is extremely difficult to determine whether certain features of complex fluids can be realized in thermal equilibrium or if they inherently require a non-equilibrium driving force.
This problem of distinguishing equilibrium and non-equilibrium features is a fundamental and unavoidable question when trying to understand complex systems.

Among complex fluids found in nature, the cytoplasm is one of the most intensively studied systems in the fields of biophysics and biochemistry~\cite{Banani2017Biomolecular}.
Experimental studies have revealed that various intracellular molecules, often proteins and ribonucleic acids (RNA), can spontaneously form droplets within the cytoplasm under specific conditions~\cite{Wright1999Intrinsically,Uversky2000Why,Dunker2001Intrinsically,Tompa2002Intrinsically,Wells2008Structure,Li2012Phase,Uversky2015Intrinsically,Hyman2011Beyond,Hyman2014Liquid-liquid}.
These droplets are interpreted as non–membrane-bound organelles and are believed to regulate the spatial organization of chemical species, similarly to membrane-bound organelles.
Many types of exotic droplets have been discovered in cells to date, and these phenomena are now collectively referred to as liquid-liquid phase separation.

Inspired by these experimental findings, a number of theoretical studies have been conducted to elucidate the physical mechanisms underlying liquid–liquid phase separation~\cite{Lee2013Spatial,Shelest2024Phase,Vweza2021Liquid-liquid,Zwicker2022The,Zwicker2024Chemically}.
These studies have mainly employed coarse-grained continuum models, which have successfully explained several experimental observations.
For example, when chemical reactions are appropriately coupled to a phase-separating fluid, Ostwald ripening can be suppressed, thereby stabilizing a state in which two centrosomes coexist within a single cell~\cite{Zwicker2014Centrosomes}.
Centrosomes are considered to be an example of so-called chemically active droplets~\cite{Zwicker2024Chemically}, where non-equilibrium reaction dynamics are essential to circumvent the normal phase separation observed in equilibrium systems and to realize exotic liquid structures characterized by heterogeneous density profiles with finite length scales.

Despite these successes, the approaches adopted in previous studies face a fundamental challenge in advancing toward a deeper understanding of complex fluids.
These coarse-grained models obscure the roles of microscopic DOF, which vary widely in both structure and interaction.
To construct a coarse-grained model, significant simplifications of the thermodynamic and dynamical properties of the system are typically made; otherwise, the resulting equations of motion, usually nonlinear partial differential equations, become too complex to solve either analytically or numerically.
As a result, all microscopic properties of complex fluids such as the cytoplasm are absorbed into a limited set of parameters.
While this simplification facilitates analysis, it comes at the cost of losing much of the underlying microscopic detail.
In particular, due to the thermodynamic simplicity of coarse-grained models, it is usually assumed, rather than demonstrated, that the non-equilibrium dynamics of the system gives rise to the observed liquid structures.
However, as discussed above, this assumption cannot be taken for granted in complex fluids, and one must carefully examine which properties of the system actually require continuous energy input.

With these considerations in mind, we propose a model that better captures the microscopic DOF without directly resorting to coarse-grained models.
In this paper, we first discuss how the microscopic DOF of complex fluids should be treated in statistical mechanics from a general perspective.
This leads us to two key assumptions that allow the derivation of a class of Hamiltonians approximating complex systems.
One assumption concerns the classification of the DOF within a single molecule, and the other involves a type of mean-field approximation for relatively unimportant DOF in the system.
These two assumptions have been adopted, either explicitly or implicitly, in many previous studies, and we examine the implications of this class of Hamiltonians in thermal equilibrium.
We then construct a simple model based on this general framework and analyze its equilibrium states.
We find that the system can exhibit phase-separated states with finite length scales even in equilibrium, resembling the non-equilibrium steady states reported in earlier studies~\cite{Ohta1986Equilibrium,Glotzer1994Monte,Glotzer1994Self-consistent,Glotzer1995Reaction,Lefever1995Comment,Glotzer1995Glotzer,Toxvaerd1996Molecular,Choksi2009On}.
The key point is that each molecule possesses a number of ``hidden'' degrees of freedom.
This result motivates a reconsideration of active droplets~\cite{Zwicker2024Chemically}, which is commonly believed to depend strongly on non-equilibrium chemical reactions.

This paper is organized as follows.
In Sec.~\ref{sec:hidden}, we discuss how to approximate microscopic DOF in a complex fluid and construct a general model based on two main assumptions.
In Sec.~\ref{sec:2+2--state model}, we analyze one of the simplest examples of the model constructed in the previous section.
In Sec.~\ref{sec:summary}, we conclude this paper with a discussion and a summary.

\section{Hidden degrees of freedom}\label{sec:hidden}

\subsection{Hamiltonian}

In this section, we will present a general framework for approximating the microscopic DOF of complex fluids.
We will often refer to the cytoplasm as a representative example of complex fluids, but our analysis applies to a broader class of systems.
We will introduce two conditions that the Hamiltonian must satisfy in order to be consistent with the diversity of DOF and their interactions.
Let us consider a system of $N$ particles in $d$-dimensional space, which is in thermal equilibrium at temperature $T = 1/\beta$.
The volume of the system is denoted by $V$.
We will explain the two main assumptions in turn.

First, we classify the DOF of a single molecule.
Generally, not all DOF directly contribute to the formation of liquid structure.
For example, liquid-liquid phase separation in cells is controlled by so-called intrinsically disordered regions of proteins~\cite{Wright1999Intrinsically,Uversky2000Why,Dunker2001Intrinsically,Tompa2002Intrinsically,Wells2008Structure,Li2012Phase,Uversky2015Intrinsically,Hyman2011Beyond,Hyman2014Liquid-liquid}, meaning that the phase separation is induced only by a part of the DOF of proteins and the other DOF have less impact on the spatial structures of the cytoplasm.
This observation leads us to assume the following condition:
\begin{assumption}[Separability of DOF]\label{asm:1}
    Each particle has two types of DOF $\bs{\sigma}$ and $\bs{\tau}$ in addition to the position $\bs{r}$ and the momentum $\bs{\pi}$.
    The former are called \emph{external} DOF and directly interact with the other particles through the pair interaction potential $\psi_{\bs{\sigma},\bs{\sigma'}} (\bs{r}, \bs{r}')$.
    The latter are called \emph{internal} DOF and only appear in the single-particle potential energy $\varepsilon (\bs{r}, \bs{\sigma}, \bs{\tau})$.
\end{assumption}
The external DOF $\bs{\sigma}$ directly affect structural properties of the system through the pair potential and can induce phase separation.
Thus $\bs{\sigma}$ include intrinsically disordered regions of a protein in the above example.
In contrast, the internal DOF $\bs{\tau}$ represent the remaining part of a molecule.
As a result, the total Hamiltonian of this system is given by
\eq{
    \mathcal H_0 \qty ( \qty { \bs{\pi}_i, \bs{r}_i, \bs{\sigma}_i, \bs{\tau}_i }_{i=1}^N ) & \coloneqq \mathcal K \qty ( \qty { \bs{\pi}_i }_{i=1}^N ) + \Psi \qty ( \qty { \bs{r}_i, \bs{\sigma}_i }_{i=1}^N ) + \sum_{i=1}^N \varepsilon \qty (\bs{r}_i, \bs{\sigma}_i, \bs{\tau}_i) \label{Hamiltonian 0}
}
where $\bullet_i$ is the observable $\bullet$ of particle $i$, $\mathcal K$ is the kinetic energy term, and 
\eq{
    \Psi \qty ( \qty { \bs{r}_i, \bs{\sigma}_i }_{i=1}^N ) & \coloneqq \sum_{i>j} \psi_{\bs{\sigma}_i,\bs{\sigma}_j} \qty (\bs{r}_i, \bs{r}_j) .
}

Two comments are in order.
First, if a particle is not spherical, the position $\bs{r}$ is interpreted as the position of its center of mass and the other relative coordinates measured from the center of mass are absorbed into the external DOF $\bs{\sigma}_{\mr{rel}}$.
Accordingly, the momentum $\bs{\pi}$ is the generalized momentum including the relative momentum conjugate to $\bs{\sigma}_{\mr{rel}}$ in addition to the center-of-mass momentum $\bs{p}$.
As a result, the term $\mathcal K$ is generally a complicated function of $\bs{\pi}$ though it does not affect the structural properties of equilibrium liquids in classical statistical mechanics.

Second, it is possible to treat different types of molecules using the Hamiltonian in Eq.~\eqref{Hamiltonian 0}.
Suppose, for example, that the system has two types of molecules: a monomer \ce{A} and a dimer \ce{A2}.
We model the former as a hard sphere and the latter as a hard dumbbell, in which two hard spheres are connected by a stiff rod.
The DOF of a dimer are the position $\bs{r}$ of its center of mass, the angle\footnote{$\bs{\theta}$ is a $d-1$-dimensional vector in the $d$-dimensional system.} $\bs{\theta}$ of the rod, and their conjugate momenta while the monomers do not have $\bs{\theta}$ and the angular momentum.
In this case, one can define $\bs{\sigma}$ and $\varepsilon$ as follows:
\eq{
    \bs{\sigma} & = ( \kappa, \bs{\theta} ) , \label{sigma dimer} \\
    \varepsilon \qty ( \bs{r}, \bs{\sigma} ) & = 
    \begin{cases}
        \varepsilon_{\ce{A}} \qty ( \bs{r} ) & (\kappa = \ce{A},\ \bs{\theta} = 0) \\
        \infty & (\kappa = \ce{A},\ \bs{\theta} \neq 0) \\
        \varepsilon_{\ce{A2}} \qty ( \bs{r}, \bs{\theta} ) & (\kappa = \ce{A2}) 
    \end{cases} , \label{epsilon dimer}
}
where, for this example, we assumed that there is no internal DOF for simplicity.
Likewise, it is possible to treat different types of molecules by introducing an additional discrete DOF, $\kappa$, representing the type.
In particular, this framework is applicable to the setup of Ref.~\cite{Laha2024Chemical}, in which only a single type of particles called a scaffold controls phase separation.

Let us introduce the other assumption, which is an approximation of the DOF that are less significant than $\bs{\sigma}$ or $\bs{\tau}$.
Before introducing the general statement of this assumption, we consider a simple example in a previous study.
In Ref.~\cite{Kamimura2024Thermodynamic}, a system with a growing volume was analyzed to describe the cell growth thermodynamically.
In real cells, the volume is determined by a membrane called the phospholipid bilayer and molecules in this membrane interact with the system of interest, i.e., cytoplasm.
Thus to model the membrane at the completely microscopic scale, we need to incorporate phospholipid molecules into the model and consider the interaction between them and the other molecules.
However, in Ref.~\cite{Kamimura2024Thermodynamic}, the system is simply surrounded by a moving hard wall because the realistic features of phospholipids do not matter and the only essential property is that the system volume is variable in time.
In this case, the interaction between phospholipids and the other molecules is approximated by a single-particle potential energy
\eq{
    \varepsilon_{\mr{wall}} (\bs{r}, \bs{\sigma}, \bs{\tau}) = 
    \begin{cases}
        0 & (\bs{r} \in \mathcal V_t) \\
        \infty & (\bs{r} \in \mathcal V_t)
    \end{cases} \label{wall}
    ,
}
where $\mathcal V_t$ denotes the region accessible to the particles and $t$ denotes the time.
The cell growth is represented by the time-dependence of $\mathcal V_t$ and its boundary $\partial \mathcal V_t$ represents the membrane.
As illustrated in this example, there are some DOF that can be approximated by time-dependent, or fluctuating, effective parameters in the single-particle potential energy because it is usually unnecessary to treat all DOF at the completely microscopic scale.
This can be viewed as a mean-field approximation of intracellular DOF.
Based on this consideration, we reach the following condition: 
\begin{assumption}[Mean-field approximation]\label{asm:2}
    The single-particle potential energy $\varepsilon$ depends on a fluctuating variable $\bs{\lambda}_t (\bs{r}, \bs{\sigma}, \bs{\tau})$ called the \emph{underlying} DOF; namely, $\varepsilon$ is rewritten as a functional of $\bs{\lambda}_t$
    \eq{
    \varepsilon \qty ( \bs{r}, \bs{\sigma}, \bs{\tau} ) \xrightarrow{} \varepsilon \qty [ \bs{r}, \bs{\sigma},\bs{\tau} | \bs{\lambda}_t ] .
    }
    We collectively refer to the internal and underlying DOF as the \emph{hidden} DOF and the system with the same external DOF, but without the hidden DOF, is called the \emph{original} system.
\end{assumption}
If $\bs{\lambda}_t$ has a sinusoidal dependence on the time $t$, for example, the system is driven out of equilibrium.
In contrast, if the total system is in thermal equilibrium including the fluctuations of $\bs{\lambda}_t$, the free energy is minimized with respect to $\bs{\lambda}_t$.
In the example in Eq.~\eqref{wall}, if the free energy is minimized with respect to the system volume, we obtain the isobaric ensemble.
This is the situation that we will assume in this paper and thus Assumption~\ref{asm:2} is modified as
\begin{manualassumption}{2'}[Mean-field approximation in equilibrium]\label{asm:2'}
    The single-particle potential energy $\varepsilon$ depends on a time-independent variable $\bs{\lambda} (\bs{r}, \bs{\sigma}, \bs{\tau})$ called the \emph{underlying} DOF; namely, $\varepsilon$ is rewritten as a functional of $\bs{\lambda}$
    \eq{
    \varepsilon \qty ( \bs{r}, \bs{\sigma}, \bs{\tau} ) \xrightarrow{} \varepsilon \qty [ \bs{r}, \bs{\sigma},\bs{\tau} | \bs{\lambda} ] .
    }
    The system is in thermal equilibrium including the underlying DOF, meaning that the free energy is minimized with respect to $\bs{\lambda}$, or equivalently, $\bs{\lambda}$ in the partition function is integrated out.
\end{manualassumption}
Using Assumptions~\ref{asm:1} and \ref{asm:2'}, the Hamiltonian of our model is given by
\eq{
    \mathcal H \qty [ \qty { \bs{\pi}_i, \bs{r}_i, \bs{\sigma}_i, \bs{\tau}_i }_{i=1}^N \bigg| \bs{\lambda} ] & \coloneqq \mathcal K \qty ( \qty { \bs{\pi} }_{i=1}^N ) + \Psi \qty ( \qty { \bs{r}_i, \bs{\sigma}_i }_{i=1}^N ) + \sum_{i=1}^N \varepsilon \qty [\bs{r}_i, \bs{\sigma}_i, \bs{\tau}_i | \bs{\lambda}] \label{Hamiltonian} ,
}
which leads to the grand canonical partition function
\eq{
    \Xi_{\beta} \qty [ \varepsilon[ \bs{r}, \bs{\sigma}, \bs{\tau} | \bs{\lambda} ] ] & \coloneqq \int \DD \bs{\lambda} \sum_{N=0}^\infty \frac{1}{N!} \qty ( \prod_{i=1}^N \bigintsss_{\substack{\bs{\pi}_i \in \mathbb{R}^{d_\pi} \\ \bs{r}_i \in \mathbb{R}^d\\ \bs{\sigma}_i \in \mathcal X_\sigma \\ \bs{\tau}_i \in \mathcal X_\tau}} \frac{\dd \bs{\pi}_i\dd \bs{r}_i \dd \bs{\sigma}_i \dd \bs{\tau}_i}{(2\pi)^d} ) \exp \qty ( -\beta \mathcal H \qty [ \qty { \bs{\pi}_i, \bs{r}_i, \bs{\sigma}_i, \bs{\tau}_i }_{i=1}^N \bigg| \bs{\lambda} ] ) , \label{Xi def}
}
where $d_\pi$ is the dimension of the space of the generalized momentum and $\mathcal X_{\sigma,\tau}$ is the state spaces for $\bs{\sigma}$ and $\bs{\tau}$\footnote{
The components of both $\bs{\sigma}$ and $\bs{\tau}$ can be either discrete or continuous.
The integral $\int_{\bs{\sigma} \in \mathcal X_\sigma, \bs{\tau} \in \mathcal X_\tau} \dd \bs{\sigma} \dd \bs{\tau}$ needs to be interpreted as a summation for discrete DOF.}.
Note that the chemical potential is absorbed into $\varepsilon$ and thus not written explicitly.
For notational simplicity, we will not write the domain of the integration in Eq.~\eqref{Xi def} from the next subsection.

Finally, we make a brief comment on chemical reactions.
Since we focus on thermal equilibrium, we do not need to consider chemical reactions explicitly~\cite{Landau1980Statistical}.
The chemical reactions only govern the conservation laws of numbers of molecules in equilibrium, which only affect the choice of the statistical ensemble.
Thus after computing the grand partition function in Eq.~\eqref{Xi def} and the associated grand potential, we can simply perform a Legendre transform to impose the conservation laws consistent with the chemical reactions of our interest.
For example, in the case of the model defined in Eqs.~\eqref{sigma dimer} and \eqref{epsilon dimer}, the number of atoms of element \ce{A} is conserved if the system is closed and a reaction \ce{2A <=> A2} occurs.
As will be explained in the next subsection, however, the conservation laws are implemented by appropriately choosing $\bs{\lambda}$ and thus our model can be used for any set of chemical reactions in equilibrium.
In this sense, the field $\bs{\lambda}$ is viewed as ``generalized stoichiometry'' in the context of chemical thermodynamics.

\subsection{Free energy}

Here we compute the grand potential of the general system introduced in the previous subsection.
Note that we will ignore the terms that are negligible in the thermodynamic limit without explicit mention.
Let us first define the free energy of the internal DOF conditioned on $(\bs{r}, \bs{\sigma})$, i.e., 
\eq{
    \beta \nu \qty [\bs{r}, \bs{\sigma} | \bs{\lambda}] & \coloneqq - \log \int \dd \bs{\tau} e^{- \beta \varepsilon \qty [ \bs{r}, \bs{\sigma}, \bs{\tau} | \bs{\lambda} ] } \label{nu} .
}
This is interpreted as the spatially varying chemical potential~\cite[Ch.~3]{Hansen2013Theory}.
Also for convenience of explanation, we introduce the grand potential $\tilde \Phi_\beta$ with $\bs{\lambda}$ fixed
\eq{
    \beta \tilde \Phi_\beta \qty [ \nu \qty [\bs{r}, \bs{\sigma} | \bs{\lambda}] ] & \coloneqq - \log \sum_{N=0}^\infty \frac{1}{N!} \bigintsss \qty ( \prod_{i=1}^N \frac{\dd \bs{\pi}_i\dd \bs{r}_i \dd \bs{\sigma}_i \dd \bs{\tau}_i}{(2\pi)^d} ) \exp \qty ( -\beta \mathcal H \qty [ \qty { \bs{\pi}_i, \bs{r}_i, \bs{\sigma}_i, \bs{\tau}_i }_{i=1}^N \bigg| \bs{\lambda} ] ) \notag \\[2pt]
    & = - \log \sum_{N=0}^\infty \frac{1}{N!} \bigintsss \qty ( \prod_{i=1}^N \frac{\dd \bs{\pi}_i\dd \bs{r}_i \dd \bs{\sigma}_i}{(2\pi)^d} e^{-\beta \nu \qty[ \bs{r}_i, \bs{\sigma}_i | \bs{\lambda} ]} ) e^{ -\beta \mathcal K \qty ( \qty {\bs{\pi}_i}_{i=1}^N ) - \beta \Psi \qty ( \qty { \bs{r}_i, \bs{\sigma}_i }_{i=1}^N ) } ,
}
where in the second line we used the definition of $\nu$ in Eq.~\eqref{nu}.
Using $\tilde \Phi_\beta$, Eq.~\eqref{Xi def} is written as
\eq{
    \Xi_{\beta} \qty [ \varepsilon \qty [ \bs{r}, \bs{\sigma}, \bs{\tau} | \bs{\lambda} ] ] & = \int \DD \bs{\lambda}\, e^{ - \beta \tilde \Phi_\beta \qty [ \nu \qty [\bs{r}, \bs{\sigma} | \bs{\lambda}] ] } \label{Xi with tilde Phi}
}
Following the standard procedure of classical density functional theory~\cite[Ch.~3]{Hansen2013Theory}, we rewrite $\tilde \Phi_\beta$ as a Legendre transform with respect to $\nu$
\eq{
    \tilde \Phi_\beta \qty [ \nu \qty [\bs{r}, \bs{\sigma} | \bs{\lambda} ] ] & = \min_{\rho^{\mr{ex}} (\bs{r}, \bs{\sigma})} \qty ( \beta \mathcal F_{\beta} \qty [ \rho^{\mr{ex}}(\bs{r},\bs{\sigma}) ] + \int \dd \bs{r} \dd \bs{\sigma} \rho^{\mr{ex}} (\bs{r},\bs{\sigma}) \beta \nu \qty [\bs{r}, \bs{\sigma} | \bs{\lambda}] ) , \label{DFT}
}
where $\rho^{\mr{ex}}$ is the Legendre dual of $\nu$ and interpreted as the density field, and $\mathcal F_\beta$ denotes the Helmholtz free energy of the original system as a functional of $\rho^{\mr{ex}}$.
The Legendre transform in Eq.~\eqref{DFT} is the starting point of classical density functional theory and $\mathcal F_\beta$ is usually called the density functional~\cite[Ch.~3]{Hansen2013Theory}\footnote{In standard textbooks on classical density functional theory such as Ref.~\cite[Ch.~3]{Hansen2013Theory}, only single-component systems are discussed, i.e., there is no additional DOF like $\bs{\sigma}$ and $\bs{\tau}$.
However, the multi-component systems can be treated in the same way as single-component ones as explained in Ref.~\cite[Sec.~6]{Morita1961A}.}.
To proceed, we perform the Laplace approximation, or the saddle-point approximation, with respect to $\lambda$ in Eq.~\eqref{Xi with tilde Phi} and have
\eq{
    \Xi_{\beta} \qty [ \varepsilon \qty [ \bs{r}, \bs{\sigma}, \bs{\tau} | \bs{\lambda} ] ] & = \exp \qty ( - \min_{\bs{\lambda} (\bs{r},\bs{\sigma},\bs{\tau})} \beta \tilde \Phi_\beta \qty [ \nu \qty [ \bs{r}, \bs{\sigma} | \bs{\lambda} ] ] ) \notag \\
    & = \exp \qty ( - \min_{\rho^{\mr{ex}}(\bs{r},\bs{\sigma}), \bs{\lambda} (\bs{r},\bs{\sigma},\bs{\tau})} \qty ( \beta \mathcal F_{\beta} \qty [ \rho^{\mr{ex}}(\bs{r},\bs{\sigma}) ] + \int \dd \bs{r} \dd \bs{\sigma} \rho^{\mr{ex}} (\bs{r},\bs{\sigma}) \beta \nu \qty [\bs{r}, \bs{\sigma} | \bs{\lambda}] ) ) , \label{Xi}
}
where we used Eq.~\eqref{DFT} in the second line.
The field $\rho^{\mr{ex}}{}^\dagger$ that minimizes Eq.~\eqref{Xi} corresponds to the average density field in equilibrium
\eq{
    \rho^{\mr{ex}}{}^\dagger (\bs{r}, \bs{\sigma}) & = \left \langle \sum_{i=1}^N \delta \qty ( \bs{r} - \bs{r}_i ) \delta \qty ( \bs{\sigma} - \bs{\sigma}_i ) \right \rangle_{\Xi_\beta} ,
}
where $\delta$ is the delta function and $\langle \bullet \rangle_{\Xi_\beta}$ denotes the average over the grand canonical ensemble defined by $\Xi_\beta$.
As a result, we have the grand potential
\eq{
    \beta \Phi_\beta \qty [ \varepsilon \qty [ \bs{r}, \bs{\sigma}, \bs{\tau} | \bs{\lambda} ] ] & \coloneqq - \log \Xi_\beta \qty [ \varepsilon \qty [ \bs{r}, \bs{\sigma}, \bs{\tau} | \bs{\lambda} ] ] \notag \\
    & = \min_{\rho^{\mr{ex}}(\bs{r},\bs{\sigma}), \bs{\lambda} (\bs{r},\bs{\sigma},\bs{\tau})} \qty ( \beta \mathcal F_{\beta} \qty [ \rho^{\mr{ex}}(\bs{r},\bs{\sigma}) ] - \int \dd \bs{r} \dd \bs{\sigma} \rho^{\mr{ex}} (\bs{r},\bs{\sigma}) \beta \nu \qty [ \bs{r}, \bs{\sigma} | \bs{\lambda} ] ) . \label{Phi}
}
This minimization problem is transformed into
\eq{
    0 & = \fdv{\beta \mathcal F_\beta \qty [ \rho^{\mr{ex}} (\bs{r}, \bs{\sigma}) ]}{\rho^{\mr{ex}} (\bs{r}, \bs{\sigma})} - \beta \nu \qty [ \bs{r}, \bs{\sigma} | \bs{\lambda} ] , \label{rho} \\
    0 & = \fdv{\bs{\lambda} (\bs{r}, \bs{\sigma}, \bs{\tau})} \int \dd \bs{r}' \dd \bs{\sigma}' \rho^{\mr{ex}} (\bs{r}',\bs{\sigma}') \beta \nu \qty [ \bs{r}', \bs{\sigma}' | \bs{\lambda} ] . \label{lambda}
}

Eqs.~\eqref{Phi}--\eqref{lambda} are sufficient to discuss the thermodynamic properties of the system, but we rewrite them in two different forms which are easy to interpret.
First, we formally perform the minimization with respect to $\bs{\lambda}$ in Eq.~\eqref{Phi}. 
Since the solution $\bs{\lambda}^\dagger$ of this minimization problem is a functional of $\rho^{\mr{ex}}$, we can rewrite Eq.~\eqref{Phi} as
\eq{
    \beta \Phi_\beta \qty [ \varepsilon \qty [ \bs{r}, \bs{\sigma}, \bs{\tau} | \bs{\lambda}^\dagger ] ] & = \min_{\rho^{\mr{ex}}(\bs{r},\bs{\sigma})} \qty ( \beta \mathcal F_{\beta} \qty [ \rho^{\mr{ex}}(\bs{r},\bs{\sigma}) ] - \int \dd \bs{r} \dd \bs{\sigma} \rho^{\mr{ex}} (\bs{r},\bs{\sigma}) \beta \nu^\dagger \qty [ \bs{r}, \bs{\sigma} | \rho^{\mr{ex}} ] ) \notag \\
    & = \min_{\rho^{\mr{ex}}(\bs{r},\bs{\sigma})} \qty ( \beta \mathcal F_{\beta} \qty [ \rho^{\mr{ex}}(\bs{r},\bs{\sigma}) ] + \beta \mathcal F'_{\beta} \qty [ \rho^{\mr{ex}}(\bs{r},\bs{\sigma}) ] ) \eqqcolon \min_{\rho^{\mr{ex}}(\bs{r},\bs{\sigma})} \beta \mathcal F_\beta^{\mr{tot}} \qty [ \rho^{\mr{ex}}(\bs{r},\bs{\sigma}) ] , \label{F tot}
}
where we plugged the solution $\bs{\lambda}^\dagger$ of the minimization problem in Eq.~\eqref{Phi} into $\nu$ and wrote
\eq{
    \nu \qty [ \bs{r}, \bs{\sigma} | \bs{\lambda}^\dagger ] = \nu^\dagger \qty [ \bs{r}, \bs{\sigma} | \rho^{\mr{ex}} ] 
}
Also we set
\eq{
    \beta \mathcal F'_{\beta} \qty [ \rho^{\mr{ex}}(\bs{r},\bs{\sigma}) ] & \coloneqq - \int \dd \bs{r} \dd \bs{\sigma} \rho^{\mr{ex}} (\bs{r},\bs{\sigma}) \beta \nu^\dagger \qty [ \bs{r}, \bs{\sigma} | \rho^{\mr{ex}} ] . \label{F'}
}
The minimization problem in Eq.~\eqref{F tot} is rewritten as
\eq{
    0 & = \fdv{\beta \mathcal F_\beta \qty [ \rho^{\mr{ex}} (\bs{r}, \bs{\sigma}) ]}{\rho^{\mr{ex}} (\bs{r}, \bs{\sigma})} + \fdv{\beta \mathcal F'_\beta \qty [ \rho^{\mr{ex}} (\bs{r}, \bs{\sigma}) ]}{\rho^{\mr{ex}} (\bs{r}, \bs{\sigma})} , \label{minimization F'}
}
which is free from the hidden DOF and thus interpreted as an equation for the original system.
Also the second term, originally from the hidden DOF, can be interpreted as a driving force acting on the original system. 
In this sense, the equilibrium states of the total system are generally ``deformed'' equilibrium states of the original free energy $\mathcal F_\beta$ and the deformation results from the pseudo-driving force in the second term of Eq.~\eqref{minimization F'}.
We will give examples of this pseudo-driving force in the next section and discuss its meaning.

The second form is derived from the Legendre transform of Eq.~\eqref{nu} with respect to $\varepsilon$
\eq{
    \beta \nu \qty [\bs{r}, \bs{\sigma} | \bs{\lambda}] & = \min_{p(\bs{\tau} || \bs{r}, \bs{\sigma})} \int \dd \bs{\tau} \qty ( p(\bs{\tau} || \bs{r}, \bs{\sigma}) \log p(\bs{\tau} || \bs{r}, \bs{\sigma}) + p(\bs{\tau} || \bs{r}, \bs{\sigma}) \beta \varepsilon \qty [ \bs{r}, \bs{\sigma}, \bs{\tau} | \bs{\lambda} ] ) ,
}
where $p(\bs{\tau} || \bs{r}, \bs{\sigma})$ is the probability distribution of $\bs{\tau}$ conditioned on $\bs{r}$ and $\bs{\sigma}$.
Using this expression, the grand potential is rewritten as
\eq{
    \beta \Phi_\beta \qty [ \varepsilon \qty [ \bs{r}, \bs{\sigma}, \bs{\tau} | \bs{\lambda} ] ] & = \min_{\rho^{\mr{ex}}(\bs{r},\bs{\sigma}), p(\bs{\tau} || \bs{r}, \bs{\sigma}), \bs{\lambda} (\bs{r},\bs{\sigma},\bs{\tau})} \bigg ( \beta \mathcal F_{\beta} \qty [ \rho^{\mr{ex}}(\bs{r},\bs{\sigma}) ] \notag \\
    & \qquad \left. + \int \dd \bs{r} \dd \bs{\sigma} \dd \bs{\tau} \rho^{\mr{ex}} (\bs{r},\bs{\sigma}) \qty ( p(\bs{\tau} || \bs{r}, \bs{\sigma}) \log p(\bs{\tau} || \bs{r}, \bs{\sigma}) + p(\bs{\tau} || \bs{r}, \bs{\sigma}) \beta \varepsilon \qty [ \bs{r}, \bs{\sigma}, \bs{\tau} | \bs{\lambda} ] ) \right ) \notag \\
    & = \min_{\rho(\bs{r},\bs{\sigma},\bs{\tau}), \bs{\lambda} (\bs{r},\bs{\sigma},\bs{\tau})} \left ( \beta \mathcal F_{\beta} \qty [ \rho^{\mr{ex}}(\bs{r},\bs{\sigma}) ] - \int \dd \bs{r} \dd \bs{\sigma} \rho^{\mr{ex}} (\bs{r}, \bs{\sigma}) \log \rho^{\mr{ex}} (\bs{r}, \bs{\sigma}) \right. \notag \\
    & \qquad \left. + \int \dd \bs{r} \dd \bs{\sigma} \dd \bs{\tau} \qty ( \rho (\bs{r},\bs{\sigma},\bs{\tau}) \log \rho (\bs{r},\bs{\sigma},\bs{\tau}) + \rho (\bs{r}, \bs{\sigma}, \bs{\tau}) \beta \varepsilon \qty [ \bs{r}, \bs{\sigma}, \bs{\tau} | \bs{\lambda} ] ) \right ) , \label{Phi 2}
}
where $\rho (\bs{r}, \bs{\sigma}, \bs{\tau}) \coloneqq \rho^{\mr{ex}} (\bs{r}, \bs{\sigma}) p (\bs{\tau} || \bs{r}, \bs{\sigma})$ is the joint distribution of $(\bs{r}, \bs{\sigma}, \bs{\tau})$ and we did not explicitly write the condition $\rho^{\mr{ex}} (\bs{r}, \bs{\sigma}) = \int \dd \bs{\tau} \rho (\bs{r}, \bs{\sigma}, \bs{\tau})$ for notational simplicity.
Using the expression in Eq.~\eqref{Phi 2}, we can see that the conservation laws can be included into this free energy by modifying the definition of $\varepsilon$ as follows.
Any conservation law is expressed as 
\eq{
    \int \dd \bs{r} \dd \bs{\sigma} \dd \bs{\tau} \rho \qty ( \bs{r}, \bs{\sigma}, \bs{\tau} ) \Lambda \qty ( \bs{\sigma}, \bs{\tau} ) & = N_C = \mr{const.}, \label{conservation law}
}
where $\Lambda (\bs{\sigma}, \bs{\tau})$ is a function to define the conservation law\footnote{
There is no mathematical difficulty even if $\Lambda$ depends on the position $\bs{r}$.
In particular, the case $\Lambda(\bs{r}) = \delta (\bs{r}-\bs{r}')$ is necessary to impose the incompressibility condition in Sec.~\ref{sec:2+2--state model}.
}.
For example, when the total number of particles is conserved, we have $\Lambda (\bs{\sigma}, \bs{\tau}) = 1$.
In the case of the model defined in Eqs.~\eqref{sigma dimer} and \eqref{epsilon dimer}, we have $\Lambda (\kappa, \bs{\theta}) = \delta_{\kappa,\ce{A}} + 2 \delta_{\kappa,\ce{A2}}$ in a closed system with the reaction $\ce{2A <=> A2}$.
Thus we can introduce a Lagrange multiplier $\lambda_0$ for each conservation law and rewrite $\varepsilon$ as
\eq{
    \varepsilon_{\mr{new}} \qty [ \bs{r}, \bs{\sigma}, \bs{\tau} | \bs{\lambda}, \lambda_0 ] & \coloneqq  \varepsilon \qty [ \bs{r}, \bs{\sigma}, \bs{\tau} | \bs{\lambda} ] + \lambda_0 \Lambda \qty ( \bs{\sigma}, \bs{\tau} ) .
}
Also, we need to add a constant $-\beta \lambda_0 N_C$ to $\beta \Phi_\beta$ and the minimization problem is rewritten as
\eq{
    \beta \Phi_\beta^{\mr{new}} \qty [ \varepsilon_{\mr{new}} \qty [ \bs{r}, \bs{\sigma}, \bs{\tau} | \bs{\lambda}, \lambda_0 ] ] & = \min_{\rho(\bs{r},\bs{\sigma},\bs{\tau}), \bs{\lambda} (\bs{r},\bs{\sigma},\bs{\tau})} \left ( \beta \mathcal F_{\beta} \qty [ \rho^{\mr{ex}}(\bs{r},\bs{\sigma}) ] - \int \dd \bs{r} \dd \bs{\sigma} \rho^{\mr{ex}} (\bs{r}, \bs{\sigma}) \log \rho^{\mr{ex}} (\bs{r}, \bs{\sigma}) \right. \notag \\
    & \qquad + \int \dd \bs{r} \dd \bs{\sigma} \dd \bs{\tau} \qty ( \rho (\bs{r},\bs{\sigma},\bs{\tau}) \log \rho (\bs{r},\bs{\sigma},\bs{\tau}) + \rho (\bs{r}, \bs{\sigma}, \bs{\tau}) \beta \varepsilon \qty [ \bs{r}, \bs{\sigma}, \bs{\tau} | \bs{\lambda} ] ) \notag \\
    & \qquad \left. + \beta \lambda_0 \qty ( \int \dd \bs{r} \dd \bs{\sigma} \dd \bs{\tau} \rho \qty ( \bs{r}, \bs{\sigma}, \bs{\tau} ) \Lambda \qty ( \bs{\sigma}, \bs{\tau} ) - N_C ) \right ) .
}
Since the value of $\lambda_0$ is determined to satisfy Eq.~\eqref{conservation law}, the last line of this equation eventually vanishes, which gives us the free energy with the desired conservation law.

To conclude this section, we consider the case where there are no underlying DOF $\bs{\lambda}$.
In this case, Eq.~\eqref{lambda} is unnecessary and the problem simplifies to
\eq{
    0 & = \fdv{\beta \mathcal F_\beta \qty [ \rho^{\mr{ex}} (\bs{r}, \bs{\sigma}) ]}{\rho^{\mr{ex}} (\bs{r}, \bs{\sigma})} - \beta \nu \qty [ \bs{r}, \bs{\sigma} ] .
}
The first term represents the chemical potential of the original system and the second term corresponds to the effective field due to the internal DOF $\bs{\tau}$.
One can see that there is a hierarchical relationship between $\bs{\sigma}$ and $\bs{\tau}$; the latter affects the former through the effective field, but not vice versa.
The system is equivalent to the original one subject to the external field $\nu$.
In contrast, when the system has the underlying DOF $\bs{\lambda}$, the two types of DOF, $\bs{\sigma}$ and $\bs{\tau}$, interact with each other, and the second term in Eq.~\eqref{rho} depends on $\rho^{\mr{ex}}$ through $\bs{\lambda}^\dagger$ as shown in Eq.~\eqref{minimization F'}, which leads to non-trivial equilibrium states that are absent in the original system as will be shown in the next section.

\setcounter{equation}{0}
\section{2+2--state model}\label{sec:2+2--state model}

Here we consider a simple, but important, example of the class of Hamiltonians introduced in Sec.~\ref{sec:hidden}:
\eq{
    \bs{\sigma} & = \sigma , \\
    \bs{\tau} & = \tau , \\
    \mathcal X_\sigma & = \qty { \sigma | \sigma = \ce{A}, \ce{B} } , \\
    \mathcal X_\tau & = \qty { \tau | \tau = \ce{A}, \ce{B} } , \\
    \varepsilon \qty [ \bs{r}, \bs{\sigma}, \bs{\tau} | \bs{\lambda} ] & = \qty ( u_\sigma + u_\tau \hat{\mathcal T} (\bs{r}) ) \lambda (\bs{r}) ,
}
where $\hat{\mathcal T}$ is an operator that can be integrated by parts, or self-adjoint
\eq{
    \int \dd \bs{r} A(\bs{r}) \hat{\mathcal T} (\bs{r}) B(\bs{r}) & = \int \dd \bs{r} B(\bs{r}) \hat{\mathcal T} (\bs{r}) A(\bs{r}) .
}
The constant $u_{\sigma,\tau}$ is given by
\eq{
    \begin{cases}
    u_{\ce{A}} = + 1 , \\
    u_{\ce{B}} = - 1 .
    \end{cases}
}
We call this model the 2+2--state model.
This is one of the minimal models that belong to the class introduced in Sec.~\ref{sec:hidden}, i.e., it has only a single external DOF and a single internal DOF, each of which is composed of two descrete states, and they are coupled linearly with the underlying DOF $\lambda$.
To further simplify the problem, we assume the incompressibility condition\footnote{One can easily impose this type of condition in the same way as conservation laws, as discussed in the previous subsection.}
\eq{
    \rho^{\mr{tot}} & = \rho_{\ce{A}} (\bs{r}) + \rho_{\ce{B}} (\bs{r}) = \mr{const.} \label{incompress}
}
Then the remaining unknown variables are $\Delta \rho^{\mr{ex}} (\bs{r}) \coloneqq \rho_{\ce{A}}^{\mr{ex}} (\bs{r}) - \rho_{\ce{B}}^{\mr{ex}} (\bs{r})$ and $\lambda (\bs{r})$.
Also for notational simplicity we rewrite all field variables as
\eq{
    A(\bs{r}, \bs{\sigma}, \bs{\tau}) = A_{\sigma\tau} (\bs{r}).
}
In this setup, Eq.~\eqref{rho} is rewritten as
\eq{
    0 & = \fdv{\beta \mathcal F_{\beta} \qty [ \Delta \rho^{\mr{ex}} \qty ( \bs{r} ) ]}{\Delta \rho^{\mr{ex}} (\bs{r})} + \beta \lambda (\bs{r}) , \label{rho 2+2 AC}
}
and from Eq.~\eqref{lambda} we have
\eq{
    0 & = \sum_{\sigma\tau} \beta \qty ( u_\sigma + u_\tau \hat{\mathcal T} (\bs{r}) ) \qty ( \rho^{\mr{ex}}_{\sigma} (\bs{r}) \frac{e^{-\beta u_\tau \hat{\mathcal T} (\bs{r}) \lambda (\bs{r})}}{\sum_{\tau'} e^{-\beta u_{\tau'} \hat{\mathcal T} (\bs{r}) \lambda (\bs{r})}} ) \notag \\
    & = \sum_\sigma \beta u_\sigma \rho_\sigma^{\mr{ex}} (\bs{r}) + \sum_\tau \beta u_\tau \hat{\mathcal T} (\bs{r}) \frac{e^{-\beta u_\tau \hat{\mathcal T} (\bs{r}) \lambda (\bs{r})}}{\sum_{\tau'} e^{-\beta u_{\tau'} \hat{\mathcal T} (\bs{r}) \lambda (\bs{r})}} \sum_\sigma \rho_\sigma^{\mr{ex}} (\bs{r}) \notag \\
    & = \beta \Delta \rho^{\mr{ex}} (\bs{r}) - \rho^{\mr{tot}} \beta \hat{\mathcal T} (\bs{r}) \tanh \beta \hat{\mathcal T} (\bs{r}) \lambda (\bs{r}) , \label{tanh} 
}
which leads to
\eq{
    \beta \lambda (\bs{r}) & = \hat{\mathcal T}^{-1} (\bs{r}) \operatorname{arctanh} \qty ( \frac{1}{\rho^{\mr{tot}}} \hat{\mathcal T}^{-1} (\bs{r}) \Delta \rho^{\mr{ex}} (\bs{r}) ) 
}
with
\eq{
    \qty | \frac{1}{\rho^{\mr{tot}}} \hat{\mathcal T}^{-1} (\bs{r}) \Delta \rho^{\mr{ex}} (\bs{r}) | < 1 . \label{ineq}
}
As a result, Eq.~\eqref{rho 2+2 AC} reduces to
\eq{
    0 & = \fdv{\beta \mathcal F_{\beta} \qty [ \Delta \rho^{\mr{ex}} \qty (\bs{r}) ]}{\Delta \rho^{\mr{ex}} (\bs{r})} + \hat{\mathcal T}^{-1} (\bs{r}) \operatorname{arctanh} \qty ( \frac{1}{\rho^{\mr{tot}}} \hat{\mathcal T}^{-1} (\bs{r}) \Delta \rho^{\mr{ex}} (\bs{r}) ) \notag \\
    & = \fdv{\beta \mathcal F_{\beta} \qty [ \Delta \rho^{\mr{ex}} \qty (\bs{r}) ]}{\Delta \rho^{\mr{ex}} (\bs{r})} + \fdv{\beta \mathcal F_\beta' \qty [ \Delta \rho^{\mr{ex}} (\bs{r}) ]}{\Delta \rho^{\mr{ex}} (\bs{r})} . \label{2+2 equation}
}
where
\eq{
    \beta \mathcal F_\beta' \qty [ A (\bs{r}) ] & = \beta \mathcal F_\beta^{\mr{free}} \qty [ (1/\rho^{\mr{tot}}) \hat{\mathcal T}^{-1} \qty (\bs{r}) A (\bs{r}) ] , \\
    \beta \mathcal F_\beta^{\mr{free}} \qty [ A (\bs{r}) ] & = \rho^{\mr{tot}} \int \dd \bs{r} \qty [ \frac{1 + A (\bs{r})}{2} \log \frac{1 + A (\bs{r})}{2} + \frac{1 - A (\bs{r})}{2} \log \frac{1 - A (\bs{r})}{2} ] \label{F free main} .
}
Note that $\mathcal F_\beta'$ was first defined in Eq.~\eqref{F'} and that $\mathcal F^{\mr{free}}$ is the free energy of a collection of free, or non-interacting, spins, see the Appendix.
From Eq.~\eqref{F tot}, Eq.~\eqref{2+2 equation} is equivalent to the minimization problem 
\eq{
    \beta \Phi_\beta \qty [ \varepsilon_{\sigma\tau} [\bs{r} | \bs{\lambda}] ] & = \min_{\Delta \rho^{\mr{ex}} (\bs{r})} \qty ( \beta \mathcal F_{\beta} \qty [ \Delta \rho^{\mr{ex}} \qty (\bs{r}) ] + \beta \mathcal F_\beta' \qty [ \Delta \rho^{\mr{ex}} (\bs{r}) ] ) \notag \\
    & = \min_{\substack{\Delta \rho^{\mr{ex}} (\bs{r}), \Delta \rho^{\mr{aux}} (\bs{r})\\\Delta \rho^{\mr{aux}} \qty (\bs{r}) = (1/\rho^{\mr{tot}}) \hat{\mathcal T}^{-1} \qty (\bs{r}) \Delta \rho^{\mr{ex}} (\bs{r})}} \qty ( \beta \mathcal F_{\beta} \qty [ \Delta \rho^{\mr{ex}} \qty (\bs{r}) ] + \beta \mathcal F_\beta^{\mr{free}} \qty [ \Delta \rho^{\mr{aux}} \qty (\bs{r}) ] ) , \label{2+2 minimization}
}
where we introduced the auxiliary field $\Delta \rho^{\mr{aux}}$.
Finally if $\rho^{\mr{tot}} \gg \Delta \rho^{\mr{ex}} (\bs{r})$, the second term in Eq.~\eqref{2+2 equation} can be expanded and we have
\eq{
    0 & = \fdv{\beta \mathcal F_{\beta} \qty [ \Delta \rho^{\mr{ex}} \qty ( \bs{r} ) ]}{\Delta \rho^{\mr{ex}} (\bs{r})} + \frac{1}{\rho^{\mr{tot}}} 
    \qty ( \hat{\mathcal T}^{-1} (\bs{r}) )^2 \Delta \rho^{\mr{ex}} (\bs{r}) , \label{expanded}
}
This form will also be useful later.
Below we will analyze two cases of $\hat{\mathcal T}$ and discuss the equilibrium density profiles in both cases.

\subsection{Allen-Cahn-type equilibrium states}\label{sec:Allen-Cahn-type}

\subsubsection{Theory}

Assume that $\hat{\mathcal T}$ is a constant.
We have from Eq.~\eqref{2+2 minimization} and \eqref{expanded}
\eq{
    \beta \Phi_\beta \qty [ \varepsilon_{\sigma\tau} [\bs{r} | \bs{\lambda}] ] & = \min_{\substack{\Delta \rho^{\mr{ex}} (\bs{r}), \Delta \rho^{\mr{aux}} (\bs{r})\\ \sqrt{\rho^{\mr{tot}}} \Delta \rho^{\mr{aux}} \qty (\bs{r}) = \alpha \Delta \rho^{\mr{ex}} (\bs{r})}} \qty ( \beta \mathcal F_{\beta} \qty [ \Delta \rho^{\mr{ex}} \qty (\bs{r}) ] + \beta \mathcal F_\beta^{\mr{free}} \qty [ \Delta \rho^{\mr{aux}} \qty (\bs{r}) ] ) , \\[2pt]
    0 & = \fdv{\beta \mathcal F_{\beta} \qty [ \Delta \rho^{\mr{ex}} \qty ( \bs{r} ) ]}{\Delta \rho^{\mr{ex}} (\bs{r})} + \alpha^2 \Delta \rho^{\mr{ex}} (\bs{r}) , \label{Allen-Cahn}
}
where $\alpha \coloneqq \sqrt{1/\rho^{\mr{tot}}} \hat{\mathcal T}^{-1}$.
The expression in Eq.~\eqref{Allen-Cahn} is equivalent to the stationary limit of the Allen-Cahn equation with a linear reaction term
\eq{
    \pdv{\phi(\bs{r},t)}{t} & = - \Gamma \fdv{\beta \mathfrak F [\phi (\bs{r},t)]}{\phi(\bs{r},t)} - k \phi(\bs{r},t) \label{reactive AC}
}
with the following correspondence:
\eq{
    \lim_{t\to\infty} \phi(\bs{r},t) & = \Delta \rho^{\mr{ex}} (\bs{r}) , \\
    \mathfrak F & = \mathcal F_\beta , \\
    \frac{k}{\Gamma} & = \alpha ,
}
where $\phi$ is the field of the order parameter, $\mathfrak F$ is the free energy as a functional of the order parameter, $\Gamma$ is the mobility, and $k$ is the reaction rate.
Thus Eq.~\eqref{reactive AC} can be interpreted as an algorithm to sample equilibrium states of this version of the 2+2--state model.
The equivalence between Eqs.~\eqref{Allen-Cahn} and \eqref{reactive AC} does not mean that the natural dynamics of this version of the 2+2--state model is described by Eq.~\eqref{reactive AC}.
It holds only between the stationary limit of Eq.~\eqref{reactive AC} and the equilibrium states described by Eq.~\eqref{Allen-Cahn}.

Also one can easily see that $\mathcal F_\beta^{\mr{tot}} = \mathcal F_\beta + \mathcal F_\beta'$ is strongly convex, meaning the absence of a continuous phase transition.
To see this, we compute the second derivative of $\mathcal F_\beta^{\mr{tot}}$
\eq{
    \frac{\delta^2 \beta \mathcal F_\beta^{\mr{tot}}[A(\bs{r})]}{\delta A(\bs{r}_1) \delta A(\bs{r}_2)} & = \frac{\delta^2 \beta \mathcal F_{\beta} \qty [ A \qty (\bs{r}) ]}{\delta A(\bs{r}_1) \delta A(\bs{r}_2)} + \frac{\alpha^2 \delta \qty ( \bs{r}_1 - \bs{r}_2 )}{1 - \qty ( A (\bs{r}) )^2} .
}
For a sufficiently small perturbation $a(\bs{r})$, the Rayleigh quotient is given by
\eq{
    \int \dd \bs{r}_1 \dd \bs{r}_2 \frac{\delta^2 \beta \mathcal F_\beta^{\mr{tot}}[A(\bs{r})]}{\delta A(\bs{r}_1) \delta A(\bs{r}_2)} a(\bs{r}_1) a(\bs{r}_2) & = \int \dd \bs{r}_1 \dd \bs{r}_2 \frac{\delta^2 \beta \mathcal F_\beta[A(\bs{r})]}{\delta A(\bs{r}_1) \delta A(\bs{r}_2)} a(\bs{r}_1) a(\bs{r}_2) + \int \dd \bs{r} \frac{\alpha^2 a(\bs{r})^2}{1 - \qty ( A (\bs{r}) )^2} \geq \alpha^2 I > 0 ,
}
where $I \coloneqq \int\dd\bs{r} a(\bs{r})^2 > 0$ and the inequality follows from the convexity of the free energy $\mathcal F_\beta$ and $|A(\bs{r})|<1$ in Eq.~\eqref{ineq}.
As a result, $\mathcal F_\beta^{\mr{tot}} \qty [ A (\bs{r}) ]$ is strongly convex as long as $\alpha \neq 0$, meaning that the compressibility is always finite.
Thus the system has no instability associated with a continuous phase transition even if the original system described by the free energy $\mathcal F_\beta$ shows a phase transition, i.e., becomes not strongly convex.

\begin{figure}
    \centering
    \includegraphics{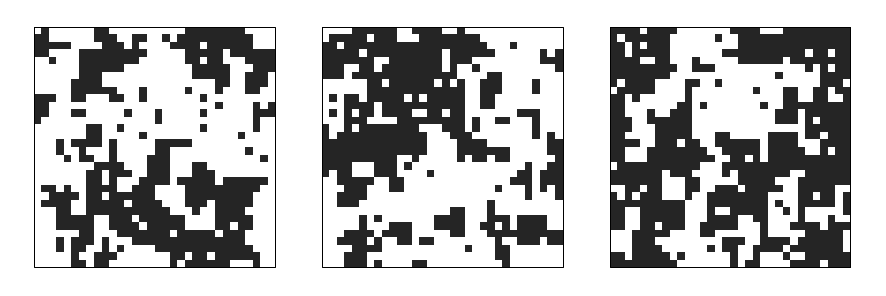}
    \caption{
    Three independent configurations of the reactive model during the Monte Carlo simulation.
    The black and white sites represent up and down spins, respectively.
    The system is in thermal equilibrium at $T = 1.0$.
    The system size is $L=32$ and the reaction probability is $p=0.1$.
    }
    \label{fig:snap}
\end{figure}

\subsubsection{Simulation}\label{sec:AC simulation}

We performed Monte Carlo simulations of this version of the 2+2--state model, which will be called the ``reactive model'' only in this subsection for convenience.
To this end, we utilized the expression in Eq.~\eqref{Allen-Cahn} and replaced the original free energy $\mathcal F_\beta$ by that of the binary lattice-gas model, or equivalently, the ferromagnetic Ising model $\mathcal F_\beta^{\mr{Ising}}$, and replace the density field $\Delta \rho^{\mr{ex}}$ by the magnetization field $m (\bs{r})$.
For simplicity, we adopted the two-dimensional square-lattice model with $N = L\times L$ spins under periodic boundary conditions.
The Hamiltonian of the standard ferromagnetic Ising model is given by
\eq{
    \mathcal H_{\mr{Ising}} \qty [ \qty { s_i }_{i=1}^N ] = - J \sum_{\langle ij\rangle} s_i s_j , \label{Ising}
}
where $s_i = \pm1$ is the spin of site $i$ and $J>0$ is the exchange coupling constant.
The sum runs over all nearest neighbor pairs of sites $i$ and $j$.
In the Monte Carlo simulation of the reactive model, each standard metropolis step of Eq.~\eqref{Ising} is followed by a ``reaction'' step in which a spin is randomly chosen and flipped with probability $p$.
This probability controls the frequency of the linear reaction described by the second term of Eq.~\eqref{reactive AC} and thus corresponds to the reaction rate $k \sim p$.

We started from a random configuration at infinite temperature and equilibrated it using the Hamiltonian of the Ising model in Eq.~\eqref{Ising}, i.e., without reaction steps, at a sufficiently high temperature $T = 5.0$ with $10^6N$ Monte Carlo steps.
The configuration at $T=5.0$ was cooled down and equilibrated to $T=0.5$ in steps of $\Delta T = 0.1$ with the same number of Monte Carlo steps at each $T$, and thus we obtained $46$ equilibrium configurations of the Ising model at different temperatures from $T=5.0$ to 0.5.
We measured some fundamental observables of these Ising configurations to compare with the reactive model.
Then we performed the simulations of the reactive model starting from the Ising configuration at each temperature.
As in the case of the Ising model, we equilibrated the system with $10^6N$ Monte Carlo steps and measured the same observables.

\begin{figure}
    \centering
    \includegraphics{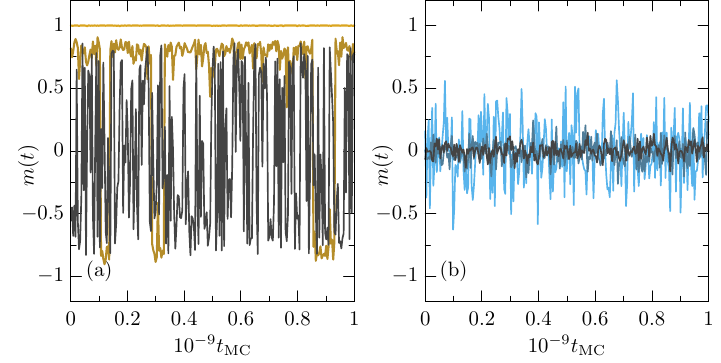}
    \caption{
    Time evolution of the magnetization per spin in (a) the ferromagnetic Ising model and (b) the reactive model.
    In panel (a), the data at three different temperatures are shown: $T=2.3$, 2.2, and 1.0 from black to yellow.
    In panel (b), the data at three different reaction probabilities are shown: $p=0.5$, 0.2, and 0.1 from black to blue.
    In both panels, the system size is set to $L=32$.
    In panel (b), the temperature is set to $T=1.0$
    }
    \label{fig:time}
\end{figure}

Fig.~\ref{fig:snap} shows three independent configurations of the reactive model during the Monte Carlo simulation.
They are in thermal equilibrium at $T = 1.0$, which is much lower than the critical temperature of the corresponding Ising model in Eq.~\eqref{Ising}, $T_c = 2/\log \qty ( 1 + \sqrt{2} ) \approx 2.269$.
The system size is $L=32$ and the reaction probability is $p=0.1$.
If $p>0$, or $\alpha > 0$, the system does not show a phase transition as discussed in the previous subsection.
However, one can see that the system shows spatial heterogeneity; namely, the correlations of black or white sites do not decay at the microscopic scale, but persist with a finite length scale.
This behavior will be analyzed in detail below.

Fig.~\ref{fig:time} shows the time evolution of the magnetization per spin
\eq{
    m (t) \coloneqq \frac{1}{N} \sum_{i=1}^N s_i (t) .
}
Panels (a) and (b) show the results of the ferromagnetic Ising model and the reactive model, respectively.
The system size is fixed to $L=32$.
Note that the time means the Monte Carlo step $t_{\mr{MC}}$.
In panel (a), the data at three different temperatures are shown: $T=2.3$, 2.2, and 1.0 from black to yellow.
The data show large fluctuations at $T=2.3$ and 2.2, which are close to $T_c$.
In particular, the magnetization at $T=2.2$ jumps between the two stable values, approximately $\pm 0.8$, which is a characteristic of a ferromagnetic state near the critical temperature.
If the temperature is sufficiently low, $T=1.0$, the magnetization is almost constant near the maximum value $m = 1.0$.
In panel (b), the data at three different reaction probabilities are shown: $p=0.5$, 0.2, and 0.1 from black to blue and the temperature is set to $T=1.0$.
In contrast to the Ising model, the reactive model does not show a phase transition and the magnetization fluctuates around $m = 0$.
However, one can see that the fluctuations grow with decreasing $p$.
This reflects the fact that the case of $p=0$ corresponds to the standard Ising model while the system does not show a phase transition as long as $p>0$.
In this sense, the limit $p\to0$ is singular.

\begin{figure}[t]
    \centering
    \includegraphics{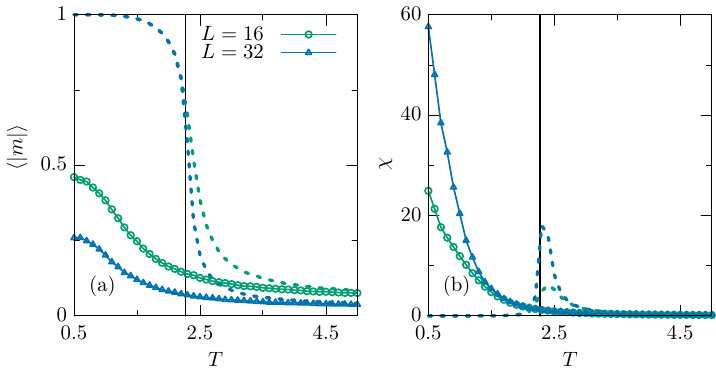}
    \caption{
    (a) Average magnetization and (b) magnetic susceptibility as functions of the temperature.
    In both panels, the solid lines with symbols show the data of the reactive model and the dotted lines show the ones of the corresponding Ising model.
    The value of $p$ is fixed to 0.2.
    The vertical black lines show the position of the critical temperature of the two-dimensional square-lattice Ising model $T_c = 2/\log \qty ( 1 + \sqrt{2} ) \approx 2.269$.
    }
    \label{fig:thermo}
\end{figure}

To examine the temperature dependence of the two systems, we define the magnetization averaged over $M=10000$ samples
\eq{
    \left\langle |m| \right\rangle & \coloneqq \frac{1}{M} \sum_{\alpha=1}^M \left | m^\alpha \right |, \label{magnetization} \\
    m^\alpha & \coloneqq \frac{1}{N} \sum_{i=1}^N m_i^\alpha ,
}
where $\alpha$ is the sample label, and the magnetic susceptibility
\eq{
    \chi & \coloneqq \beta N \qty ( \left\langle m^2 \right\rangle - \left\langle |m| \right\rangle^2 ) .
}
Fig.~\ref{fig:thermo} (a) and (b) show the average magnetization and the magnetic susceptibility, respectively, as functions of the temperature for two different system sizes $L=16$ and 32.
In both panels, the solid lines with symbols show the data of the reactive model and the dotted lines show the ones of the corresponding Ising model.
The vertical black lines show the position of the critical temperature $T_c$.
The data of the Ising model show typical behavior of the continuous phase transition.
Across the critical temperature, the jump of the magnetization, or the order parameter, becomes sharper as the system size increases.
The magnetization does not depend on the system size below the critical temperature, or in the ferromagnetic phase, while it converges to zero above the critical temperature, or the paramagnetic phase.
The susceptibility develops a peak at the critical temperature, which eventually diverges in the thermodynamic limit.
In contrast, the magnetization of the reactive model decays to zero as the system size increases at any temperatures, meaning that the absence of a continuous phase transition as discussed above.
However, the susceptibility grows rapidly in the limit of zero temperature, indicating the existence of a growing length scale.

\begin{figure}
    \centering
    \includegraphics{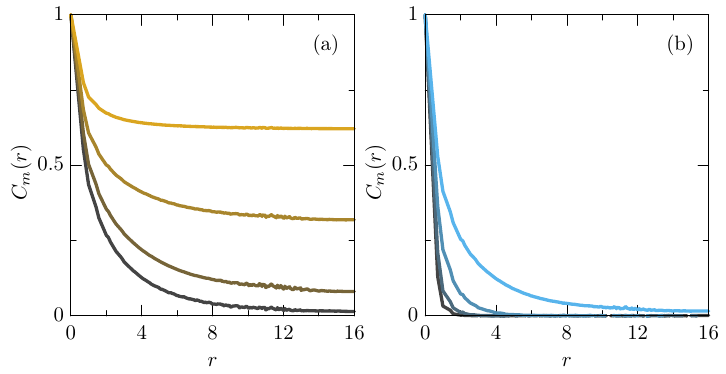}
    \caption{
    Correlation functions in (a) the ferromagnetic Ising model and (b) the reactive model.
    In panel (a), the data at four different temperatures are shown: $T=2.5$, 2.4, and 2.3 from black to yellow.
    In panel (b), the data at four different values of $p$ are shown: $p=1.0$, 0.5, 0.2, and 0.1 from black to blue.
    In both panels, the system size is set to $L=32$.
    In panel (b), the temperature is set to $T=1.0$
    }
    \label{fig:corr}
\end{figure}

The growth of $\chi$ at low temperatures suggests some spatial anomaly in the magnetization profile and thus to directly analyze the spatial structure of the reactive model, we computed the correlation function of the magnetization field
\eq{
    C_m (r) & \coloneqq \left \langle \frac{1}{N} \sum_{\substack{i\geq j=1\\|\bs{r}_i-\bs{r}_j|=r}}^N s_i s_j \right \rangle ,
}
where $\bs{r}_i$ is the position of site $i$.
Fig.~\ref{fig:corr} shows the correlation functions of (a) the Ising model and (b) the reactive model.
In panel (a), the data at four different temperatures are shown: $T=2.5$, 2.4, 2.3, and 2.2 from black to yellow.
In panel (b), the data at four different reaction probabilities are shown: $p=1.0$, 0.5, 0.2, and 0.1 from black to blue.
In both panels, the system size is set to $L=32$ and in panel (b) the temperature is set to $T=1.0$.
The correlation function of the Ising model again show typical behavior: it does not decay completely in the ferromagnetic phase because of the spontaneous magnetization.
On the other hand, the correlation function of the reactive model always decays to zero, which is consistent with the absence of a phase transition, but it develops a slowly decaying tail in the limit $p\to0$ at sufficiently low temperatures.
Thus the system shows spatial heterogeneity of the magnetization field with a finite length scale $\xi(T, p)$ as observed in Fig.~\ref{fig:snap}.
In the case of the standard Ising model, the temperature needs to be fine-tuned around $T_c$ to observe spatial heterogeneity with the correlation length $\xi(T)$.
In contrast, since the true phase transition is suppressed by the hidden DOF in the reactive model, the heterogeneous density profile is easily observed by simply decreasing $p$ and $T$.

\subsection{Cahn-Hillard-type equilibrium states}

\subsubsection{Theory}

In this subsection, we consider a slightly different version of the 2+2--state model to relate our theory to previous studies on models that exhibit microphase separation.
When $\hat{\mathcal T}^2 = -(\rho^{\mr{tot}}/\alpha')\nabla^2$, we have from Eq.~\eqref{expanded}
\eq{
    0 & = \nabla^2 \fdv{\beta \mathcal F_{\beta} \qty [ \Delta \rho^{\mr{ex}} \qty ( \bs{r} ) ]}{\Delta \rho^{\mr{ex}} (\bs{r})} - \alpha' \Delta \rho^{\mr{ex}} (\bs{r}) , \label{Cahn-Hilliard}
}
which is equivalent to the stationary limit of the Cahn-Hillard equation with a linear reaction term~\cite{Glotzer1994Monte,Glotzer1994Self-consistent,Glotzer1995Reaction,Lefever1995Comment,Glotzer1995Glotzer}
\eq{
    \pdv{\phi(\bs{r},t)}{t} & = \Gamma \nabla^2 \fdv{\beta \mathfrak F [\phi (\bs{r},t)]}{\phi(\bs{r},t)} - k \phi(\bs{r},t) \label{reactive CH} ,
}
with the following correspondence:
\eq{
    \lim_{t\to\infty} \phi(\bs{r},t) & = \Delta \rho^{\mr{ex}} (\bs{r}) , \\
    \mathfrak F & = \mathcal F_\beta , \\
    \frac{k}{\Gamma} & = \alpha' . 
}
Thus we can interpret Eq.~\eqref{reactive CH} as an algorithm to sample equilibrium states of this version of the 2+2--state model similarly to the case of Sec.~\ref{sec:Allen-Cahn-type}.

Note that assuming $\hat{\mathcal T}^2 = -(\rho^{\mr{tot}}/\alpha')\nabla^2$ is equivalent to requiring
\eq{
    \Delta \rho^{\mr{in}} (\bs{r}) = \sqrt{-\qty (\rho^{\mr{tot}}/\alpha')^2\nabla^2} \Delta \rho^{\mr{ex}} (\bs{r}) . \label{fractional Laplacian}
}
The partial differential equations involving the fractional Laplacian $(-\nabla^2)^a$ with $0<a<1$ appear in various fields of physics and mathematics~\cite{Lischke2020What}.
For example, anomalous diffusion in fractal systems is often described by the diffusion equation with the Laplacian replaced by the fractional one.
Thus it might be the case that the relation in Eq.~\eqref{fractional Laplacian} naturally emerges if the system has a fractal structure, e.g., networks or porous media, which is often observed in living systems.

\subsubsection{Comparison with previous studies}

There are many theoretical and numerical studies on Eq.~\eqref{reactive CH}~\cite{Ohta1986Equilibrium,Glotzer1994Monte,Glotzer1994Self-consistent,Glotzer1995Reaction,Lefever1995Comment,Glotzer1995Glotzer,Toxvaerd1996Molecular,Choksi2009On} and a comparison of our results with those gives some insights into our model.
Before explaining the complications that arise when physically interpreting Eq.~\eqref{reactive CH}, we summarize some of its mathematical aspects regarding the stationary states, which were extensively studied in Ref.~\cite{Choksi2009On}.
In the following discussion, we always assume that the original system described by $\mathcal F_\beta$, or equivalently $\mathfrak F$, shows a phase separation at sufficiently low temperatures.
The most crucial point is that the stationary states of the model in Eq.~\eqref{reactive CH} do not show a global phase separation as long as $k>0$.
At sufficiently low temperatures, the uniform density field is destabilized and the system shows a microphase separation.
It was numerically confirmed that the resulting stationary density field $\lim_{t\to\infty} \phi (\bs{r}, t)$ is heterogeneous with various patterns depending on the values of the model parameters, though a mathematically rigorous proof of the stability of each pattern is challenging.
In any case, the length scale of each pattern remains finite and does not diverge.
Based on these mathematical or numerical results, it is obvious that Eq.~\eqref{Cahn-Hilliard} also does not show a global phase separation and instead shows a microphase separation depending on the values of $\beta$, $\alpha'$ and the other material parameters.

We then present a brief overview of the research on Eq.~\eqref{reactive CH}, particularly focusing on its physical meaning.
This model was first proposed as a coarse-grained dynamical model of diblock copolymer melts~\cite{Ohta1986Equilibrium}.
In this context, the second term in the right-hand side of Eq.~\eqref{reactive CH} results from the variation of the free energy
\eq{
    \beta \mathfrak F^{\mr{poly}} [ \phi (\bs{r}, t) ] & \coloneqq - k \int \dd \bs{r}_1 \dd \bs{r}_2 G \qty (\bs{r}_1 - \bs{r}_2) \phi (\bs{r}_1, t) \phi (\bs{r}_2, t) ,
}
where $G$ denotes Green's function of the Laplacian.
This contribution to the free energy represents the long-range interaction of polymers.
Using the total free energy $\mathfrak F + \mathfrak F^{\mr{poly}}$, Eq.~\eqref{reactive CH} was derived as a standard Cahn-Hillard equation
\eq{
    \pdv{\phi(\bs{r},t)}{t} & = \nabla^2 \fdv{\phi(\bs{r},t)} \qty ( \Gamma \beta \mathfrak F [\phi (\bs{r},t)] + \beta \mathfrak F^{\mr{poly}} [\phi (\bs{r},t)] )
}
to describe the \emph{equilibrium} and \emph{conservative} dynamics of polymer melts; namely, the system is not driven by an external field and each polymer does not change its chemical composition.
In contrast, it was later pointed out that the model in Eq.~\eqref{reactive CH} can be used to describe the \emph{non-equilibrium} and \emph{non-conservative} dynamics of a unimolecular chemical reaction
\eq{
    \ce{A <=> B} \label{unimolecular}
}
in a system which exhibits a phase separation into the \ce{A}-rich and \ce{B}-rich phases~\cite{Glotzer1994Monte,Glotzer1994Self-consistent,Glotzer1995Reaction,Lefever1995Comment,Glotzer1995Glotzer}.
Under this interpretation, the first term in Eq.~\eqref{reactive CH} corresponds to the normal phase-separation kinetics of the non-reactive system and the second one corresponds to the externally-driven chemical reaction in Eq.~\eqref{unimolecular}.
Note that this system cannot reach an equilibrium state as discussed in Ref.~\cite{Lefever1995Comment,Glotzer1995Glotzer}.
After these pioneering works~\cite{Glotzer1994Monte,Glotzer1994Self-consistent,Glotzer1995Reaction,Lefever1995Comment,Glotzer1995Glotzer}, a number of similar nonlinear partial differential equations have been proposed in the context of liquid-liquid phase separation observed in cells, see, e.g., Ref.~\cite{Weber2019Physics,Zwicker2022The,Zwicker2024Chemically}.
These equations of motion are interpreted as describing the non-equilibrium dynamics of chemical reactions during phase separation in the same spirit as the work in Ref.~\cite{Glotzer1994Monte,Glotzer1994Self-consistent,Glotzer1995Reaction,Lefever1995Comment,Glotzer1995Glotzer}.

In this study, however, the expression in Eq.~\eqref{Cahn-Hilliard} was derived under different assumptions compared with these previous studies.
Eq.~\eqref{Cahn-Hilliard} results directly from the variational principle of thermodynamics and describes equilibrium states of a version of the 2+2--state model.
In this model, the number of particles with $\sigma = \ce{A}$ or \ce{B} is not conserved and in this sense Eq.~\eqref{Cahn-Hilliard} describes \emph{equilibrium} states of a \emph{non-conservative} system.
Although the original system is a binary liquid mixture, the whole system shows a heterogeneous density profile with a microphase separation even in equilibrium, i.e., without energy injection, because of the hidden DOF and their interactions.
This situation is similar to the version in Sec.~\ref{sec:Allen-Cahn-type} though it does not show a microphase separation.

\section{Summary and conclusion}\label{sec:summary}

Many theoretical studies have been conducted to describe the exotic structures of complex fluids.
In particular, liquid-liquid phase separation in cells has recently attracted significant interest in biophysics and biochemistry and a number of models have been proposed to explain the observed droplet formation.
However, studies based on conventional continuum descriptions have limitations because these models obscure the distinct roles of the microscopic DOF in complex fluids.
Also, it remains a challenge to describe such multi-component systems even in thermal equilibrium.
To address these issues and deepen our understanding of the thermodynamic properties of complex fluids, we adopt two main approximations for the DOF and construct a general Hamiltonian based on them.
The first assumption is the separability of DOF of a single molecule, meaning that some DOF of a molecule, called internal DOF, do not interact directly with the other molecules in the system.
The second assumption is a mean-field approximation for relatively unimportant DOF.
Since it is usually unnecessary to model all DOF at a fully microscopic level, these less important DOF can be coarse-grained, resulting in time-dependent or fluctuating effective parameters in the single-particle potential energy.
If the system reaches equilibrium while including these coarse-grained DOF, the resulting free energy is minimized with respect to the corresponding effective parameters as well.
From these two assumptions, the system contains “hidden” DOF, which include the internal DOF and the effective parameters arising from coarse-graining.
These hidden DOF affect the system only through the single-particle potential and do not directly influence the interaction potential.
As a result, the grand potential of the system consists of two terms:
the first corresponds to the free energy of the system without hidden DOF, and the second arises from the hidden DOF.
These two contributions can be interpreted as the free energy of the original system modified by a pseudo-driving force introduced by the hidden DOF.

We then showed, using a simple model, that a system with such hidden DOF can exhibit a heterogeneous density profile with a finite length scale.
While such heterogeneous states have been investigated particularly in the context of liquid-liquid phase separation in cells with a strong focus on their non-equilibrium aspects, our results indicate that some of these states can be realized as equilibrium states under reasonable assumptions for modeling the microscopic DOF of complex fluids.
From a biological perspective, this may explain why droplets are ubiquitous in living systems.
These droplets are believed to play important roles in regulating biological functions~\cite{Zwicker2014Centrosomes}, and in this sense, droplet formation is considered evolutionarily advantageous.
However, droplets can also form spontaneously, without continuous energy input, when intracellular molecules become highly complex and acquire many distinct DOF, similar to the hidden DOF considered in this study. 
In other words, liquid-liquid phase separation could be a by-product of protein evolution.
We again emphasize the importance of determining whether a given phenomenon truly requires energy input or can be partially realized under equilibrium conditions.

Finally, we briefly discuss future perspectives, particularly focusing on the dynamics of our model.
To fully understand complex fluids in nature, it is essential to consider their dynamical properties.
When analyzing the dynamics of our model, whether in equilibrium or out of equilibrium, one of the most challenging problems is how to model chemical reactions, or transitions between different values of the DOF, in a way that is consistent with both thermodynamics and quantum mechanics.
Since it is both practically impossible and typically unnecessary to analyze a realistic quantum Hamiltonian directly in order to explain macroscopic fluid structure, chemical reactions must be modeled as classical stochastic processes that are microscopically compatible with quantum dynamics and macroscopically consistent with the thermodynamic properties discussed in this paper.
The conventional approach is based on transition state theory~\cite{Hanggi1990Reaction}, in which chemical reactions are assumed to proceed via activation processes. 
It should be worthwhile to investigate how hidden DOF affect the activation energy.

\section*{Acknowledgments}

We are grateful to the members of the Laboratory for Quantitative Biology and to the participants of the meeting supported by Toyota Konpon Research Institute, Inc. (Yanagisawa Group) for their fruitful discussions.
MS also thanks A. Ikeda and N. Oyama for their insightful comments.
This work was supported by JSPS KAKENHI Grant Numbers 25K17353 and 25H01365, JST CREST Grant Number JPMJCR2011, and the Initiative on Promotion of Supercomputing for Young or Women Researchers, Supercomputing Division, Information Technology Center, The University of Tokyo.

\section*{Author declarations}

\subsection*{Conflict of interest}

There are no conflicts to declare.

\section*{Data availability}

The data that support the findings of this study are available from the corresponding author upon reasonable request.

\appendix*

\renewcommand{\theequation}{A.\arabic{equation}}
\section{Statistical mechanics of free spins}\label{sec:free}

Here we compute the free energy of a collection of $N$ free, or non-interacting, spins $\{s_i\}_{i=1}^N$, where $s_i = \pm 1$.
Each spin is subject to the magnetic field $h_i$ and the Hamiltonian reads
\eq{
    \mathcal H_{\mr{free}} \qty [ \qty {s_i, h_i}_{i=1}^N ] & \coloneqq -\sum_{i=1}^N s_i h_i .
}
The partition function of this system is given by
\eq{
    Z_\beta^{\mr{free}} \qty [ \qty {h_i}_{i=1}^N ] & \coloneqq \sum_{s_1=\pm1} \cdots \sum_{s_N=\pm1} e^{-\beta \mathcal H_{\mr{free}} \qty [ \qty {s_i, h_i}_{i=1}^N ]} = \prod_{i=1}^N \qty ( e^{\beta h_i} + e^{-\beta h_i} ) ,
}
which leads to the free energy
\eq{
    \beta G_\beta^{\mr{free}} \qty [ \qty {h_i}_{i=1}^N ] & \coloneqq - \log Z_\beta^{\mr{free}} \qty [ \qty {h_i}_{i=1}^N ] = - \sum_{i=1}^N \log \qty ( e^{\beta h_i} + e^{-\beta h_i} ) . \label{G free}
}
The $h_i$-derivative of the free energy gives the average magnetization of spin $i$
\eq{
    m_i \qty (h_i) & \coloneqq \langle s_i \rangle_{G_\beta^{\mr{free}}} = \frac{1}{\beta} \pdv{\beta G_\beta^{\mr{free}} \qty [ \qty {h_i}_{i=1}^N ]}{h_i} = \frac{e^{\beta h_i}}{e^{\beta h_i} + e^{-\beta h_i}} .
}
Using $\{m_i\}_{i=1}^N$, we perform the Legendre transform of Eq.~\eqref{G free} with respect to $\{h_i\}_{i=1}^N$
\eq{
    \beta F_\beta^{\mr{free}} \qty [ \qty {m_i}_{i=1}^N ] & = \max_{\qty {h_i}_{i=1}^N} \qty ( \beta G_\beta^{\mr{free}} \qty [ \qty {h_i}_{i=1}^N ] + \beta \sum_{i=1}^N h_i m_i ) \notag \\
    & = \sum_{i=1}^N \qty ( \frac{1+m_i}{2} \log \frac{1+m_i}{2} + \frac{1-m_i}{2} \log \frac{1-m_i}{2} ) \notag \\ 
    & = \rho^{\mr{tot}} \sum_{i=1}^N \ell^d \qty ( \frac{1+m_i}{2} \log \frac{1+m_i}{2} + \frac{1-m_i}{2} \log \frac{1-m_i}{2} ) \notag \\
    & = \rho^{\mr{tot}} \int \dd \bs{r} \qty ( \frac{1+m \qty (\bs{r})}{2} \log \frac{1+m \qty (\bs{r})}{2} + \frac{1-m \qty (\bs{r})}{2} \log \frac{1-m \qty (\bs{r})}{2} ) \eqqcolon \beta \mathcal F_\beta^{\mr{free}} \qty [ m \qty (\bs{r}) ] ,\label{F free}
}
where $\ell$ is the lattice constant, $\rho^{\mr{tot}} \coloneqq 1/\ell^d$ is the total density of spins, and $m(\bs{r})$ is the magnetization field.
Eq.~\eqref{F free} is equivalent to Eq.~\eqref{F free main} in the main text.

\end{document}